\documentclass[letterpaper,american,reprint,amsmath,amssymb,prd,showpacs]{revtex4-1}
\usepackage[latin9]{inputenc}
\usepackage{textcomp}
\usepackage{amstext}
\usepackage{amssymb}
\usepackage{esint}

\makeatletter

\pdfpageheight\paperheight
\pdfpagewidth\paperwidth

\usepackage{dcolumn}% Align table columns on decimal point
\usepackage{bm}% bold math

\usepackage{graphicx}
\makeatother

\usepackage{babel}
\begin{document}

\title{Mass in cosmological perspective}

\author{Herman Telkamp}

\affiliation{Jan van Beverwijckstraat 104, 5017JA Tilburg, The Netherlands}
\email{herman\_telkamp@hotmail.com}

\begin{abstract}
\noindent We consider the total nonlocal energy associated with a
particle at rest in the Hubble flow, i.e., the relational energy between
this particle and all connected particles within the causal horizon.
The particle, even while at rest, partakes in relative recessional
and peculiar motion of connected particles in 3 dimensions. A geometrical
argument due to Berkeley suggests that the nonlocal mass of recessional
energy associated with the particle is 3 times its Newtonian mass.
It follows that nonlocal recessional and peculiar energy of the Universe
are equal, and match Misner-Sharp energy within the apparent horizon.
Contributions of recessional and peculiar nonlocal energy are thus
shown to generate a 6 times higher level of matter energy than expected
from the Newtonian mass. Accordingly, the nonlocal energy density
of baryons is expected to be 6 times the standard local energy density
of baryons, i.e., $\Omega_{\textrm{b,eff}}=6\Omega_{\textrm{b}}$.
At $\Omega_{\textrm{b}}\sim0.0484\pm0.0017$ (Planck 2015 results)
this predicts a nonlocal baryon energy density $\Omega_{\textrm{b,eff}}\sim0.290\pm0.010$,
in agreement with observed matter density $\Omega_{\textrm{m}}\sim0.308\pm0.012$.
The effect of nonlocal mass on solar system and galactic scales is
considered.
\end{abstract}
\maketitle

\section{Introduction}

Newtonian physics has a concept of both local and nonlocal energy.
Kinetic energy is attributed to a particle, so is localizable at the
particle's position. Gravitational potential energy, on the other
hand, is mutual and shared between particles, hence, it cannot be
localized at a point. This however means that the conserved total
energy, the sum of both, is necessarily nonlocal too. One of the main
criticisms of Newton's theory from the start has been that kinetic
energy of an object is only physically meaningful if considered in
relation to other matter, ultimately the background of the ``fixed
stars.'' Thus one can argue that in Newtonian physics \textit{all}
energy is essentially nonlocal, not just gravitational energy. It
is legitimate to ask why this would be any different in general relativity,
which actually was intended to satisfy this Machian principle. The
rather artificial distinction between local and nonlocal energy becomes
even less pertinent in the homogeneous, isotropic universe, where
both can only appear as spatially constant energy densities. The question
then is whether we can recognize nonlocal components of the density
parameter $\rho$. Considering that in terms of local energy the density
of baryonic matter can only explain about $5\%$ of the required total
energy density $\rho$, the remainder (or all) can perhaps be attributed
to nonlocal energy contributions. 

Theoretical approaches to represent nonlocal energy typically involve
the use of pseudotensors (e.g., Einstein, Landau-Lifshitz, Bergmann,
Møller) or prescriptions of quasilocal energy (e.g., Misner-Sharp,
Hawking, Brown-York, Epp). Although literature is not conclusive,
studies mostly agree on zero or constant total energy of the Universe,
at least in the flat case \citep{FaraoniCooperstock2003ApJ...587..483F,LapiedraMorales2012GReGr..44..367L,BanerjeeSen1997Prama..49..609B,BrownYork1993PhRvD..47.1407B,CooperstockIsraelit1995FoPh...25..631C,Hayward1994QuasilocalPhRvD..49..831H,Johri1995GReGr..27..313J,NesterSoVargas2008PhRvD..78d4035N,Rosen1994GReGr..26..319R,Szabados2009,Xulu2000IJTP...39.1153X}.
Unfortunately these notions of nonlocal energy (e.g., a zero-energy
universe) provide no direct information about the evolution of the
density parameter $\rho$ in a way consistent with observation. There
are various indications however, apart from the mass deficit itself,
that mass in cosmological context is not necessarily the same as mass
in local context. An intrinsic reason comes from a conjecture due
to Berkeley \citep{berkeley}, which gives rise to a different particle
mass in peculiar and recessional motion, as shown hereafter. Notice
that mass associated with nonlocal energy is nonlocal too, as it depends
on distribution and motion of interacting particles (mass of binding
energy being a familiar example). Within the context of general relativity,
the notion of (Misner-Sharp) quasilocal energy suggests that the nonlocal
energy density of cosmic matter differs from the standard local energy
density, as we shall point out first. We use $c=1$ throughout. 

\section{Misner-Sharp energy}

Misner-Sharp energy represents internal energy (kinetic and potential)
of a perfect fluid contained in a sphere of arbitrary radius \citep{MisnerSharp}.
Within the apparent horizon of FLRW universes it equals (the Schwarzschild
mass) \citep{Hayward1996MisnerSharpPhRvD..53.1938H,FaraoniPhysRevD.84.024003}
\begin{equation}
E_{\textrm{MS}}=\frac{R_{\textrm{a}}}{2G},\label{eq:EMS}
\end{equation}
while the apparent horizon radius $R_{\textrm{a}}$ satisfies \citep{Rindler}
\begin{equation}
H^{2}R_{\textrm{a}}^{2}=1+\frac{8}{3}\pi G\rho_{k}R_{\textrm{a}}^{2},\label{RA}
\end{equation}
where $H$ is the Hubble parameter and $\rho_{k}$ is curvature energy
density. Energy $E_{\textrm{MS}}(=M_{\textrm{MS}}c^{2})$ exerts a
potential $\frac{3}{2}GM_{\textrm{MS}}R_{\textrm{a}}^{-1}=\frac{3}{4}$
at the center of the sphere. Introducing the density of Misner-Sharp
energy, $\rho_{E}\equiv E_{\textrm{MS}}/\frac{4}{3}\pi R_{\textrm{a}}^{3}$,
and using Eqs.(\ref{eq:EMS}) and (\ref{RA}), one obtains the energy
equation (per unit mass)
\begin{equation}
T_{\textrm{a}}=\frac{3}{4}H^{2}R_{\textrm{a}}^{2}=2\pi G(\rho_{E}+\rho_{k})R_{\textrm{a}}^{2}=E_{\textrm{a}}-V_{\textrm{a}},\label{MS energy}
\end{equation}
where, in classical terms, $T_{\textrm{a}}\equiv\frac{3}{4}H^{2}R_{\textrm{a}}^{2}$
is kinetic energy, $E_{\textrm{a}}\equiv2\pi G\rho_{E}R_{\textrm{a}}^{2}=\frac{3}{4}$
is conserved total energy and $V_{\textrm{a}}\equiv-2\pi G\rho_{k}R_{\textrm{a}}^{2}$
is curvature energy, i.e., gravitational potential energy. Since the
energies are expressed per unit mass, they can be regarded potentials. 

There are some observations to make: (a) conservation of energy seems
to hold if defined in terms of nonlocal energy. (b) Equation (\ref{MS energy})
is actually the Friedmann equation, multiplied on both sides by the
common factor $R_{\textrm{a}}^{2}$. Hence, if the Misner-Sharp formalism
is correct, then the total density in the Friedmann equation equals
$\rho=\rho_{E}+\rho_{k}$, and nothing seems to be missing. For as
far assumed local matter density $\rho_{\textrm{m}}$ is represented,
it must take the nonlocal form of $\rho$. We shall investigate this
in the next section. (c) The Misner-Sharp formalism employs comoving
coordinates \citep{MisnerSharp}. Therefore recessional speed in these
coordinates is zero, so that Misner-Sharp energy only represents peculiar
energy of the fluid. (d) Kinetic energy $T_{\textrm{a}}=\frac{3}{4}H^{2}R_{\textrm{a}}^{2}$
is, for the appearance of $H$, naturally associated with recessional
motion of matter, while Misner-Sharp energy only expresses peculiar
energy, which one may not immediately relate to the Hubble parameter.
That is, unless the two, peculiar and recessional energy, maintain
a fixed ratio. This indeed follows from both equipartition and the
relational derivation hereafter.

\section{Relational energy\label{sec:Relational-energy}}

In the relational (Machian) view \citep{sep-RelationalTheories},
energy is exclusively a mutual property \textit{between} causally
connected particles, therefore not an intrinsic property of a particle,
meaning that local energy in fact does not exist in the relational
universe. This may be understood realizing that the potential energy
of a particle of mass $m$ equals $m\varphi$, where $\varphi$ is
the cosmic potential. Without the cosmic mass-energy present, the
potential energy of the particle would vanish. A similar argument
applies to photon energy $h\nu$, where $\nu$ is the photon frequency.
A vanishing potential would redshift the photon frequency to zero.
Note that the Misner-Sharp mass within the apparent horizon indeed
equals the Schwarzschild mass. Hence the idea that particle energy
disappears in absence of other mass is not uncommon. This dependency
can be largely disregarded though in the local frame, where spacetime
is just an ``empty'' flat Minkowski background to local physics.
Accordingly, our notions and unit of inertial mass relate to peculiar
motion of an object in some particular direction, while in the relational
view this object, even when at rest in the Hubble flow, partakes in
energy exchange of recessional and peculiar motion of cosmic mass
in \textit{all} directions. In other words, the Newtonian mass and
energy of an object in peculiar motion express only part of the total
energy associated with the object. This follows directly from Berkeley's
ontological conjectures \citep{berkeley}.

George Berkeley, an early critic of Newton, noted that one can not
meaningfully attribute a position or velocity to a single (point)
particle in empty space. Consequentially, this applies to kinetic
and potential energy too, hence to both inertial and gravitational
mass. These properties can only emerge from the interaction with other
particles, and are therefore necessarily shared, mutual properties
between particles, so not localizable in a point and not intrinsic
to a particle. Berkeley continues noting that of two particles in
otherwise empty space, only their radial distance is observable. Motion
in any perpendicular direction, like with these two particles in circular
orbit of each other, is unobservable in an empty background. Therefore
(and this is crucial), motion in nonradial direction, does \textit{not}
represent energy between two point particles. This means that both
the kinetic energy $T_{ij}$ and potential energy $V_{ij}$ between
point particles $i$ and $j$ depend \textit{only} on their separation
$R_{ij}$, or time derivative thereof, as pointed out by Poincaré
and others \citep{poincare1907science,Schroedinger,Barbour}. Note
that Newtonian potential energy 
\begin{equation}
V_{ij}=-Gm_{i}m_{j}R_{ij}^{-1}\label{Vij-1}
\end{equation}
is perfectly Machian \citep{poincare1907science}. It is indeed a
mutual, frame independent property between two connected particles
and depends geometrically only on their separation. Newtonian kinetic
energy, on the contrary, is defined relative to a frame of reference,
so is not relational. Schr\"{o}dinger \citep{Schroedinger,Barbour}
reproduced Einstein's expression of the anomalous perihelion precession
from the following definition of Machian kinetic energy, 
\begin{equation}
T_{ij}={\textstyle {\displaystyle \frac{1}{2}}}\mu_{ij}\dot{R}_{ij}^{2},\label{Tij-1-1}
\end{equation}
where $\mu_{ij}$ represents the mutual mass between particles $i$
and $j$, 
\begin{equation}
\mu_{ij}\equiv\frac{V_{ij}}{\varphi_{\textrm{p}x}}.\label{muij}
\end{equation}
The effective potential $\varphi_{\textrm{p}x}$, defined hereafter,
normalizes $\mu_{ij}$ in order to match Newtonian mass and kinetic
energy in peculiar motion \citep{TelkampPhysRevD.94.043520}. Definition
(\ref{Tij-1-1}) meets the Machian requirements: kinetic energy $T_{ij}$
is mutual between two particles, is frame independent, and depends
only on the radial component of motion. The total kinetic and potential
energy associated with particle $i$ follows from summation over all
particles within the causal radius $R_{g}$ of particle $i$, i.e.,
$T_{i}=\sum_{j}T_{ij}$ and $V_{i}=\sum_{j}V_{ij}=m_{i}\varphi_{N}$,
where
\begin{equation}
\varphi_{\textrm{N}}=-2\pi G\rho R_{g}^{2}\label{phiN}
\end{equation}
is the Newtonian potential at the center of the causal sphere. 

\subsection{Nonlocal mass\label{subsec:Non-local-mass}}

Like kinetic and potential energy, the total mass $\mu_{i}$ associated
with particle $i$ is a nonlocal, distributed property. However, the
value of $\mu_{i}$ does not follow from simple addition, i.e., $\mu_{i}\neq\sum_{j}\mu_{ij}$,
as pointed out next. 

Due to the exclusively radial relationship in Eq.(\ref{Tij-1-1}),
particle $j$ only contributes to kinetic energy $T_{i}$ and mass
$\mu_{i}$ if $\dot{R}_{ij}\neq0$. Hence $\dot{R}_{ij}=0$ effectively
nullifies the contribution of particle $j$ to both $\mu_{i}$ and
the Newtonian mass $m_{i}$. This implies that only a part of the
total connected mass, and therefore only a part $\varphi_{\textrm{p}x}$
of the total Newtonian potential $\varphi_{\textrm{N}}$, contributes
to the Newtonian mass $m_{i}$ of a particle $i$ in peculiar motion
in some direction $x$. In a homogeneous, isotropic sphere, where
all particles are in random peculiar motion, this fraction is 
\begin{equation}
\xi_{\textrm{p}x}\equiv\bigl\langle\dot{R}_{ij}^{2}\bigr\rangle/\bigl\langle v_{ij}^{2}\bigr\rangle_{\textrm{p}x}=\frac{1}{3},\label{xi p}
\end{equation}
where $v_{ij}$ is the relative speed, and $\dot{R}_{ij}$ the radial
component of $v_{ij}$, so that the effective potential in peculiar
motion in an arbitrary direction $x$ is (cf. \citep{Schroedinger,TelkampPhysRevD.94.043520})
\begin{equation}
\varphi_{\textrm{p}x}={\textstyle \xi_{\textrm{p}x}\varphi_{\textrm{N}}}={\textstyle {\displaystyle \frac{1}{3}}\varphi_{\textrm{N}}}.\label{Phip-1}
\end{equation}
Likewise the mass of peculiar motion in the $x$-direction between
particle $i$ and all connected particles equals
\begin{equation}
\mu_{i}^{(\textrm{p}x)}=\xi_{\textrm{p}x}{\textstyle \sum_{j}}\:\mu_{ij}=\frac{1}{3}\,\frac{\sum_{j}V_{ij}}{{\textstyle {\displaystyle {\textstyle \frac{1}{3}}}\varphi_{\textrm{N}}}}=m_{i},\label{mui-1}
\end{equation}
as expected. Thus Newtonian mass agrees with nonlocal mass in peculiar
motion in arbitrary direction. 

\medskip{}

Different from peculiar motion, recession is purely radial motion
between all particles, i.e., $\dot{R}_{ij}=v_{ij}$, and hence
\begin{equation}
\xi_{\textrm{r}}=\bigl\langle\dot{R}_{ij}^{2}\bigr\rangle/\bigl\langle v_{ij}^{2}\bigr\rangle_{\textrm{r}}=1.\label{eq:Xi r}
\end{equation}
The kinetic energy of recession therefore balances with the full potential;
all connected particles contribute fully. Thus the effective potential
in recessional motion is 
\begin{equation}
\varphi_{\textrm{r}}=\xi_{\textrm{r}}\varphi_{\textrm{N}}=\varphi_{\textrm{N}}.\label{Phir-1}
\end{equation}
This however means that a particle in recessional motion effectively
has an effective mass 3 times larger than the Newtonian mass in peculiar
motion. It interacts with 3 times as much mass. Indeed the total mass
between particle $i$ and all connected receding particles equals
\begin{equation}
\mu_{i}^{(r)}=\xi_{\textrm{r}}{\textstyle \sum_{j}}\:\mu_{ij}=\frac{\sum_{j}V_{ij}}{{\textstyle {\displaystyle {\textstyle \frac{1}{3}}}\varphi_{\textrm{N}}}}=3m_{i}.\label{mui}
\end{equation}
This is an intriguing consequence of Berkeley's conjectures, evidently
hinting at a possible interpretation of unidentified dark matter in
the form of existing, but unrecognized, nonlocal energy components
associated with each baryonic particle. Referencing Eqs.(\ref{mui})
and (\ref{mui-1}), the total nonlocal mass associated with particle
$i$ in the homogeneous, isotropic universe follows from adding up
the components of $\mu_{i}$
\begin{equation}
\mu_{i}=\mu_{i}^{(r)}+\mu_{i}^{(px)}+\mu_{i}^{(py)}+\mu_{i}^{(pz)}=6m_{i}.\label{mui-2}
\end{equation}
Note that this follows from geometrical considerations only. Equation
(\ref{mui-2}) reflects that nonlocal energy density associated with
baryonic matter is 6 times the local energy density. A perhaps conceptually
more satisfactory way to derive this result is through actual calculation
of the recessional and peculiar energies, as follows.

\subsection{Total recessional and peculiar energy\label{subsec:Total-recessional-and} }

We consider a unit mass test particle at rest in the Hubble flow at
the position of the comoving observer. Adopting Eq.(\ref{Tij-1-1}),
integration over the causal sphere $\mathcal{V}_{g}$ yields the recessional
Machian kinetic energy $T_{\textrm{r}}$ between the test particle
and all receding mass within the causal horizon at radius $R_{g}\equiv ar_{g}$
\citep{TelkampPhysRevD.94.043520},

\begin{equation}
T_{\textrm{r}}=\intop_{\mathcal{V}_{g}}{\textstyle \frac{1}{2}}\frac{\textrm{d}\varphi_{\textrm{r}}(r,\theta,\phi)}{\frac{1}{3}\varphi_{\textrm{r}}}\,r^{2}\dot{a}^{2}=\frac{3}{4}r_{g}^{2}\dot{a}^{2}=\frac{3}{4}H^{2}R_{g}^{2}.\label{eq:T-1-1}
\end{equation}
According to Eq.(\ref{Phir-1}), the potential in recessional motion
is the total Newtonian potential $\varphi_{\textrm{r}}=\varphi_{\textrm{N}}=-2\pi G\rho R_{g}^{2}$,
where $\rho$ is total density. Hence, the equation of total recessional
energy is 

\begin{equation}
T_{\textrm{r}}=\frac{3}{4}H^{2}R_{g}^{2}=2\pi G\rho R_{g}^{2}=-\varphi_{N}.\label{Mach rec-1}
\end{equation}
This again is the Friedmann equation, but derived from Machian principle
\citep{TelkampPhysRevD.94.043520}.

Recalling that the effective potential in peculiar motion in arbitrary
direction $x$ is $\varphi_{\textrm{p}x}=\frac{1}{3}\varphi_{\textrm{r}}=\frac{1}{3}\varphi_{\textrm{N}}$,
we expect the balancing kinetic energies to maintain the same ratio,
i.e., $T_{\textrm{p}x}=\frac{1}{3}T_{\textrm{r}}$. Thus
\begin{equation}
T_{\textrm{p}x}=\frac{1}{4}H^{2}R_{g}^{2}=-\frac{1}{3}\varphi_{\textrm{N}}.\label{Tpx}
\end{equation}
Like with recessional motion, this holds for a test particle at rest
in the Hubble flow; i.e., Eq.(\ref{Tpx}) expresses the kinetic energy
due to the $x$-component of peculiar motion (on all physical scales)
of all connected particles. The total peculiar energy associated with
the test particle, summed over 3 orthogonal directions, is
\begin{equation}
T_{\textrm{p}}\equiv T_{\textrm{p}x}+T_{\textrm{p}y}+T_{\textrm{p}z}=\frac{3}{4}H^{2}R_{g}^{2}=2\pi G\rho R_{g}^{2}=-\varphi_{\textrm{N}}.\label{Mach pec-1}
\end{equation}
Hence $T_{\textrm{p}}=T_{\textrm{r}}$, in agreement with both the
equipartion theorem ($T_{\textrm{p}x}=T_{\textrm{p}y}=T_{\textrm{p}z}=T_{\textrm{r}x}=T_{\textrm{r}y}=T_{\textrm{r}z}$)
and Misner-Sharp energy (at $R_{g}=R_{\textrm{a}}$). The nonlocal
recessional and peculiar energy combined thus add to 
\begin{equation}
T\equiv T_{r}+T_{p}=\frac{3}{2}H^{2}R_{g}^{2}=4\pi G\rho R_{g}^{2}=-\varphi=-2\varphi_{\textrm{N}}.\label{Mach energy-1}
\end{equation}
Therefore 
\begin{equation}
T=6T_{px},\label{T=00003D6Tpx}
\end{equation}
consistent with Eq.(\ref{mui-2}). 

Notice that the equivalence of inertial and gravitational mass is
implicitly satisfied by all energy equations above. The equations,
expressed per unit mass, have the common form $T_{\star}=-\varphi_{\star}$.
By definition, the kinetic energy $T_{\star}$ involves inertial mass,
and the Newtonian gravitational potential $\varphi_{\star}$ involves
gravitational mass. For an arbitrary test particle with inertial mass
$m_{\textrm{I}}$ and gravitational mass $m_{\textrm{G}}$, the equation
becomes $m_{\textrm{I}}T_{\star}=-m_{\textrm{G}}\varphi_{\star}$.
Hence $m_{\textrm{I}}=m_{\textrm{G}}$.

\section{Cosmological observation of nonlocal mass}

Contributions of both recessional and peculiar nonlocal energy in
3 spatial dimensions have been shown to generate a 6 times higher
level of matter energy than expected from the Newtonian mass of cosmic
matter. This suggests an effective nonlocal baryon energy density
of the Universe of 6 times the local energy density of baryons, i.e.,
$\Omega_{\textrm{b,eff}}=6\Omega_{\textrm{b}}$. According to Planck
2015 data \citep{Planck2015}, the baryon density is $\Omega_{\textrm{b}}h^{2}\sim0.0222\text{\textpm}0.0002$.
At $h\sim0.678\text{\textpm}0.009$ this gives $\Omega_{\textrm{b}}\sim0.0484\text{\textpm}0.0017$,
while estimated matter density is $\Omega_{\textrm{m}}\sim0.308\text{\textpm}0.012$.
The factor of 6 then predicts a total nonlocal baryon energy density
$\Omega_{\textrm{b,eff}}=6\Omega_{\textrm{b}}\sim0.290\text{\textpm}0.010$,
which matches $\Omega_{\textrm{m}}$ within the $68\%$ confidence
limits given. The nonlocal mass associated with baryonic matter thus
provides interpretation to dark matter on the cosmological scale.

\section{Observation of nonlocal mass on local scales}

The above model of nonlocal energy regards the causally connected
mass of a homogeneous isotropic Universe. By Mach's principle the
only true scale of any gravitational system is the cosmological scale,
meaning that cosmic nonlocal energy acts on any scale, even while
not necessarily recognized as such. On the other hand, a specifically
local aspect of gravitational systems is inhomogeneity and the interaction
between the system's constituents. A question then is how nonlocal
mass relates to Newtonian mass and general relativistic effects in
a gravitational system, discussed as follows. 

According to the above, the total nonlocal mass associated with a
body in the homogeneous Universe equals $m_{\textrm{eff}}=6m$, where
$m$ is the Newtonian mass of the body, i.e., the part of $m_{\textrm{eff}}$
that is observed in peculiar motion. Hence, in the relational view
Newtonian mass arises from the interaction with cosmic mass. The reason
that $\frac{5}{6}$ of $m_{\textrm{eff}}$ is not locally observable
from the motion of the body itself is that the body's peculiar motion
evidently is in one direction at the time, while recessional motion
of bodies in a gravitationally bound local system is negligible or
zero. A system, like the solar system or a galaxy, may be seen as
a distribution of interacting local bodies superposed on a homogeneous
cosmic background of relatively very low density. While the huge total
amount of cosmic mass \textit{outside} the system induces the Newtonian
mass of all bodies, the interaction of bodies $i$ and $j$ \textit{inside}
the system gives rise to additional nonlocal mass $\mu_{ij}$, which,
considering Eq.(\ref{muij}), is typically extremely small compared
with the Newtonian masses involved. That is, $\mu_{ij}\lll m_{i}+m_{j}$.
Yet, the tiny effect of the corresponding relational kinetic energy
$T_{ij}=\frac{1}{2}\mu_{ij}\dot{R}_{ij}^{2}$ between the bodies is
observable on the solar system scale, for instance as the anomalous
perihelion precession, or as Lense-Thirring frame dragging. Schr\"{o}dinger
showed that the effect of relational kinetic energy between two orbiting
bodies precisely matches the general relativistic expression of the
anomalous precession \citep{Schroedinger}. Thus Schr\"{o}dinger's
model is meaningful on both the small (solar system) and the large
(cosmic) scale. 

What this suggests is that the internal, non-Newtonian, part of the
system mass arises from the local interaction of bodies within the
system, and that the kinetic energy of these local interactions appears
to be accountable for general relativistic deviations. Moreover, at
an increasing number $N$ of particles within the system, the number
of internal interactions grows as $N(N-1)$. One therefore expects
the total internal part of the system mass {[}i.e., $\sum_{i\neq j}\sum_{j}\mu_{ij}\sim N(N-1),\;i,j=1,..N${]},
to grow exponentially faster than the Newtonian mass of the system,
which grows like $\sim N$, thus giving rise to much stronger deviations
from Newtonian behavior in more massive larger systems. This may be
of interest in the study of galaxy rotation and clusters. 

\bibliographystyle{apsrev4-1}
%%\bibliography{../../BibTex/referencescentral2}


\end{document}